# DATA-CENTRIC DESIGN: INTRODUCING AN INFORMATICS DOMAIN MODEL AND CORE DATA ONTOLOGY FOR COMPUTATIONAL SYSTEMS

Paul Knowles[1], Bart Gajderowicz[2] and Keith Dugas[3]

[1]The Human Colossus Foundation, Geneva, Switzerland
[2]Mechanical and Industrial Engineering Department, University of Toronto, Toronto, Ontario, Canada
[3]Secours.ai Inc., Sheridan, Wyoming, United States

*ABSTRACT*

*The Core Data Ontology (CDO) and the Informatics Domain Model represent a transformative approach to computational systems, shifting from traditional node-centric designs to a data-centric paradigm. This paper introduces a framework where data is categorized into four modalities: objects, events, concepts, and actions. This quadrimodal structure enhances data security, semantic interoperability, and scalability across distributed data ecosystems. The CDO offers a comprehensive ontology that supports AI development, role-based access control, and multimodal data management. By focusing on the intrinsic value of data, the Informatics Domain Model redefines system architectures to prioritize data security, provenance, and auditability, addressing vulnerabilities in current models. The paper outlines the methodology for developing the CDO, explores its practical applications in fields such as AI, robotics, and legal compliance, and discusses future directions for scalable, decentralized, and interoperable data ecosystems.*

*KEYWORDS*

*Informatics model, distributed data ecosystems, cryptographic data security, semantic interoperability, ontology design*

## 1. INTRODUCTION

The landscape of computational systems continues to evolve, yet many still rely on traditional node-centric frameworks, where IP addresses and device-centric identification dominate system architectures. While foundational in developing the internet and data communication protocols, these models present significant security challenges, especially as data scales and systems become increasingly interconnected. IP-centric models often expose critical vulnerabilities, such as data breaches, misuse of geolocation information, and inadequate access control mechanisms.

The Core Data Ontology[1] (CDO) offers a solution through a transformative shift from node-centric to data-centric design. We prioritize multimodal data categorization in a data-centric model, separating data into objects, events, concepts, and actions. This separation allows us to focus on the data's structural, epistemological, and functional aspects. It ensures its security, management, and utilization across distributed systems, providing a reassuring layer of protection for your data.

---

[1] Core Data Ontology repository represented as an OWL 2 ontology: Accessed at https://github.com/THCLab/ontology

The CDO addresses several limitations in current systems, including the lack of semantic interoperability. This term refers to the ability of different systems to interpret and understand shared data consistently and meaningfully. The CDO ensures this by providing a common understanding of data across systems, enabling seamless data exchange and integration. Other limitations it addresses include insufficient cryptographic assurances (such as data integrity and authenticity) and weak mechanisms for granular role-based access control. By decoupling data from its physical network location, the CDO not only enhances security and flexibility but also allows systems to evolve and scale without compromising the integrity of the data.

From an AI perspective, the CDO also unlocks new potential. Keith Dugas emphasizes how this data-centric approach enhances multimodal AI, especially in contexts requiring robust data provenance and auditability. For instance, securing consent receipts for data used in AI training can help develop auditable, explainable AI models. Moreover, integrating object and concept layers into AI systems (such as large vision-language models) improves real-world data classification, labeling, and semantic interpretation, advancing fields like robotics and automated reasoning. The CDO enables more sophisticated imitation learning and autonomous decision-making in environments requiring real-time object-event-action reasoning.

This paper is structured as follows. Section 2 outlines the problem statement, detailing the limitations of existing node-centric models. Section 3 highlights the specific objectives of the CDO, emphasizing its advantages over traditional models. Section 4 outlines the methodology used to design and implement the CDO. Section 5 discusses the benefits of the CDO, including security and scalability. Section 6 provides an overview of the Informatics Domain Model, a conceptual framework that underpins the CDO and guides its design and implementation. Section 7 further details the core data ontology. Section 8 presents practical applications and real-world use cases of the model, including AI integration, while Section 9 discusses scalability and future directions for research. Finally, in section 10, we conclude with a summary of our methodology and proposed model.

## 2. PROBLEM STATEMENT

In the current landscape of computational systems, there are notable challenges and limitations in existing informatics models and approaches. Traditionally, data semantics has been treated as a secondary process, overshadowed by node-centric mechanics focused on securing IP addresses during data transmission. This approach often leads to schema design being an afterthought, addressed during data curation after poorly structured data has entered the system.

One of the critical issues in current informatics models relates to the cryptographic data security properties of "integrity" and "authenticity." While both properties are crucial for ensuring data accuracy across digital networks, connections, or exchanges, they are only sometimes simultaneously adhered to, resulting in security degradation. For instance, the authenticity of a data source does not guarantee the integrity of the metadata and data within the message, creating challenges in preserving the original context of messages at various interaction points in a data lifecycle.

Data accuracy, the primary standard of data quality, encompasses several essential characteristics. Authentic data lineage ensures the traceability of data origin and its movement over time. Incorruptible content guarantees that data remains unaltered from source to target. Comprehensible meaning provides the necessary information to interpret and understand the context of the data. Unfortunately, in today's digital systems, it is common to encounter uninterpretable data from known sources in their original formats. Additionally, the prevalence of unstructured data, accounting for approximately 80% of recorded data, poses further challenges. However, data tied to structured metadata becomes more accessible for artificial intelligence, machine learning, and statistical analysis, as it provides comprehensible meaning.

The semantic design assumes a primary role by shifting the focus to a data-centric model, with consensual schemes developed for common thematic purposes. Placing data semantics as a priority enables the establishment of a stable foundation for scheme standardization, semantic interoperability, and data harmonization practices through consensus. After all, securing the data source loses its significance if the receiver cannot interpret and comprehend the transmitted data, underscoring the importance of semantic design as an essential precursor to system mechanics. In contrast to the node-centric model, where identity management takes precedence, the data-centric model prioritizes data management, recognizing the fundamental value of meaning and interpretation in making data meaningful and valuable.

To illustrate the challenges faced in existing informatics models, consider the scenario of an event notary and an action tracker. In a node-centric design, the central focus lies on the IP address, acting as a central entity in the model. However, in a data-centric design, the data itself takes center stage, leading to a shift in perspective. This approach allows role-based access to core data domains, providing granular control and ensuring appropriate access to specific data elements based on user roles and responsibilities.

For example, in a data-centric model, an event notary's role might involve witnessing and certifying stemmatic events, ensuring their accurate recording within a trackable sequence. Their access would be limited to the relevant events, ensuring the integrity and traceability of logged event data. On the other hand, an action tracker's role may focus on overseeing guardable actions, with their access enabling the monitoring and evaluation of systematic actions. This remit might include assessing data life cycle operations, such as transformation or disclosure processes, to ensure compliance and safeguard against potential risks.

This granular separation of role-based access rights is a fundamental feature of the Informatics domain model. The model establishes the four core data domains—objects, events, concepts, and actions—as integral components to support different user roles' specific needs and responsibilities. In this approach, the primary emphasis is on the content of the data rather than its physical location.

## 3. OBJECTIVES

This paper introduces a comprehensive and scalable solution based on the data-centric design paradigm, addressing the challenges and limitations of current informatics models and approaches. The paper aims to include the following objectives:

1. Enable security through the Informatics domain model.

> The first objective is to demonstrate how the Informatics model redefines security by categorizing data into quadrimodal domains. The model provides a new vantage point to design secure role-based access solutions tailored to specific user roles and responsibilities by categorizing data into the four modalities of objects, events, concepts, and actions. This approach ensures granted data access based on individuals' needs and privileges, enhancing data security and minimizing the risk of unauthorized access or data breaches.

2. Develop a core data ontology based on the Informatics domain model.

The second objective is to emphasize the development of a comprehensive and granular core data ontology built upon the elements of the Informatics model. The ontology provides a labeling solution for categorizing data components within the quadrimodal framework. It enables the systematic classification of data parts based on their characteristics, whether it pertains to morphological structure (object tagging), kinematical causality (event tagging), epistemological knowledge (concept tagging), or dynamical intelligence (action tagging). The core data ontology serves as a unifying model that data architects and system designers can reference to ensure consistent knowledge representation and understanding within the evolving landscape of distributed data ecosystems.

3. Establish the Informatics domain model as a foundational reference.

The third objective is positioning the Informatics model as a standard foundational model for system designers and data architects. The paper aims to foster a shared understanding and knowledge within the industry by promoting its adoption. The Informatics model is a unifying framework that transcends traditional boundaries, allowing stakeholders to navigate the complexities of data interoperability, governance, and security in a distributed data ecosystem. It provides a comprehensive reference point that enhances communication, collaboration, and standardization across diverse systems and domains.

Overall, the paper aims to present the Informatics model as a powerful and transformative approach that addresses the challenges of current informatics models and enables organizations to effectively navigate the evolving landscape of data-centric design and distributed data ecosystems.

## 4. METHODOLOGY

The development of the Informatics domain model and its accompanying core data ontology followed a systematic and multidisciplinary approach. The research began in July 2020, initially focusing on exploring the dualism of data semantics and inputs. The model aimed to capture the machine domains and their cryptographic assurance, represented by the dual-modal structure. To ensure the accuracy and consistency of terminology, the development team relied on the rich heritage of the Oxford English Dictionary (OED), renowned for its comprehensive definitions and authoritative language references since its first publication in February 1884. By leveraging the OED as a primary point of reference, the model integrated common and well-established definitions to avoid any shoehorning of terminology, enhancing the clarity and coherence of the model (see https://zenodo.org/records/13729820 for the associated glossary terms).

The research process spanned diverse disciplines, including data science, informatics, linguistics, semantics, mathematics, philosophy, psychology, and music theory. This multidisciplinary approach ensured the Informatics model encompassed a comprehensive understanding of various domains, resulting in a granularly accurate representation. By incorporating insights from these disciplines, the model aimed to provide a robust and versatile frame of reference for data-centric design and computational systems.

The development of the core data ontology for the Informatics model commenced in April 2023, following the completion of the model itself. An o*ntology* is a formal representation of knowledge encompassing concepts, their properties, and the relationships between them. It acts as a shared vocabulary and a set of constraints to enable effective communication, understanding, and interoperability across different stakeholders, systems, and domains [1]. This unifying concept can be incorporated into thematic ontologies, enabling dynamic data-centric search capabilities across computational systems. The Informatics model provides a flexible and adaptable foundation for organizing and categorizing data in diverse domains with an ontological underpinning to enhance dynamic search across distributed data ecosystems.

No legacy models were referenced during the Informatics model's development, ensuring a neutral and unbiased approach. This approach allowed for a bottom-up modelling process, starting from a blank canvas and progressively building the model's structure and components. Once the model was complete, further research into scientific and academic papers further enhanced the validation of the model's credibility. This research provided additional confidence in the Informatics model, reinforcing its relevance and applicability in data-centric design.

By adopting a comprehensive and interdisciplinary methodology, the Informatics model and accompanying ontology aim to provide a robust and versatile framework for understanding and navigating the complexities of data-centric design and distributed data ecosystems.

## 5. BENEFITS OF USING CORE DATA ONTOLOGY

Introducing the Core Data Ontology (CDO) enhances the ability to search for data-related elements across various storage components, including repositories, data vaults, and registries. By providing a unified semantic framework, the CDO facilitates efficient retrieval, classification, and understanding of data, offering significant advantages for academic researchers and data-driven applications alike. Here's how they might benefit:

Formalizing shared concepts and vocabularies has several benefits in the adoption and interoperability of distributed systems [1]. We outline the following benefits of the proposed Core Data Ontology.

**Structured Searching**: A CDO allows for a structured and standardized method of searching within a semantic repository. Researchers can search based on well-defined categories and relationships, ensuring their queries yield targeted results.

**Enhanced Discoverability**: With a unified ontology, disparate pieces of data become linked through shared terms, relationships, and definitions. Researchers can more easily discover related information due to this interlinking, reducing the chance of overlooking important details.

**Consistency**: As everyone uses the same ontological framework, it ensures that the data is consistent and uniform. This consistency reduces confusion and aids in data interpretation.

**Increased Interoperability**: If a researcher uses multiple databases or datasets, having a consistent ontology ensures that data from different sources can be easily merged and analyzed.

**Semantic Enrichment**: CDO enables researchers to glean more profound insights by understanding the context in which data exists rather than just the data itself.

Benefits in the Absence of an Authorisation Layer include:

1. **Open Access**: Without an authorization layer, all researchers, regardless of affiliation or seniority, can freely access and search the repository. This democratization ensures that data is available to all.

2. **Unbiased Searching**: Specific data remains accessible to all researchers, ensuring a fair and comprehensive data discovery process, as it's not hidden or gated based on a researcher's role.

For Researchers Searching the Repository, the benefits include:

1. **Efficient Queries**: By leveraging the CDO, researchers can make more specific and efficient queries, saving time and energy.
2. **Cross-disciplinary Insights**: By adopting a unified ontology, researchers can uncover links to various disciplines, facilitating interdisciplinary research endeavours.

For researchers contributing to the repository, the benefits include:

1. **Standardized Contribution**: Using the CDO as a guide, researchers can ensure that the data they contribute adheres to the repository's standards, making it more valuable to other users.
2. **Data Lineage and Provenance**: Even without a robust authorization layer, a unified ontology can assist in tracking the lineage of data, ensuring that contributors receive appropriate credit for their contributions.
3. **Feedback Loop**: As more researchers contribute using the same ontology, it can evolve to become more comprehensive, benefiting the entire community.

In conclusion, while an authorization layer would add security and role-specific access, CDO benefits to academic researchers are plentiful. The structured, consistent, and enriched environment it creates is invaluable for both data consumption and contribution.

## 6. INFORMATICS MODEL OVERVIEW

The Informatics domain model introduces a comprehensive framework that redefines the approach to data-centric design and addresses the challenges of current informatics models. The core of the Informatics model is the fulcrum between four distinct domains: the morphological structure [2] of "o*bjects,*" the kinematical causality [3] of "e*vents,*" the provision of epistemological knowledge [4] through "c*oncepts,*" and the active dynamical intelligence [5] behind "a*ctions."* These domains represent the fundamental building blocks of data and provide a holistic perspective for understanding and organizing information within computational systems.

### 6.1 Informatics Domain Model

In the morphological domain, object schemas define how textual attributes contribute to the morphological structure of systemic variables, data structures, functions, and methods. This domain emphasizes the structural organization of data, ensuring precise classification and representation of data elements based on their inherent characteristics.

The kinematical domain captures data motion and causal sequence through events, providing dynamic evidence of interrelated actions. This domain enables a comprehensive understanding of how data elements evolve through sequential processes and data states, focusing on the progression, interaction, and impact of events within computational systems.

The epistemological domain focuses on the epistemic knowledge associated with data, where concept frames define how contextual terms and relationships contribute to the understanding and interpretation of data. This domain ensures a shared understanding and enables effective communication and interoperability across diverse systems and domains.

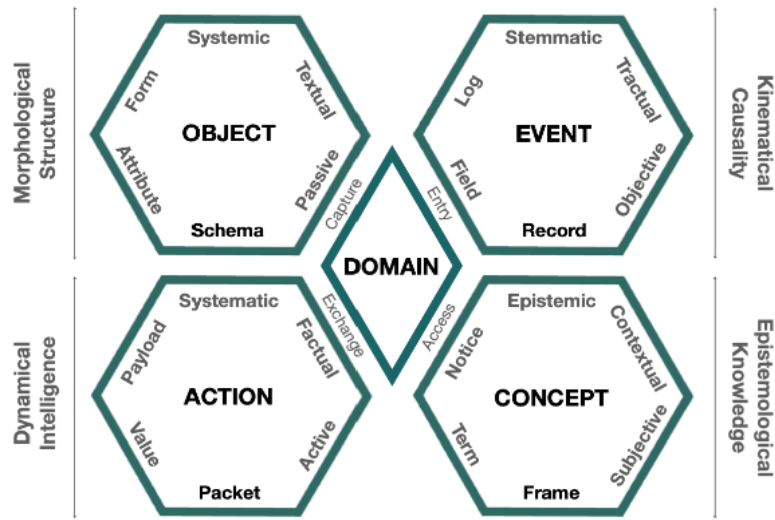

Figure 1. Informatics Domain Model.

The dynamical domain represents actionable intelligence, where action packets contain factual data to drive systematic operations in real-time. This domain encompasses the processes, decisions, and operations that leverage data to execute lifecycle tasks, focusing on dynamic decision-making and value-based actions in computational systems.

The Informatics model recognizes the interconnections and interdependencies between these domains. Objects and concepts work closely together, with objects providing structural information that concepts can use for representing entities within a subject domain. The relations from concepts represent contextual information about objects and influence the model's behavior. This correlation between objects and concepts creates a "consensual scheme," establishing consensus and agreement on the structure and interpretation of data.

Similarly, events and actions are closely intertwined. Events provide data inputs that actions can then use and evaluate. In contrast, the outputs from actions can manifest into recorded events, influencing the components' behavior after an event's completion. This correlation between events and actions creates "sovereign reason," where actions and events collaborate to drive the behavior and functionality of the computational system.

By incorporating these cross-domain correlations, the Informatics domain model enables a comprehensive and integrated approach to data-centric design. It recognizes the importance of data structure and semantic aspects, fostering a holistic understanding and facilitating efficient data management, processing, and analysis.

With a solid foundation established by the Informatics model, organizations can leverage the consensual scheme and sovereign reason aspects of data management to develop secure role-based access solutions, ensuring appropriate access to data based on user roles and responsibilities. The core data ontology based on the Informatics model provides a labelling solution that categorizes data components within the quadrimodal framework, enabling granular classification and improved search capabilities.

The Informatics model serves as a unifying framework for knowledge and understanding within the evolving landscape of distributed data ecosystems. It fosters communication, collaboration, and standardization across diverse systems and domains, establishing a foundational reference for system designers, data architects, and stakeholders.

## 6.2 CDO Functional Correlations

Constituent elements of the model can be configured into unique interactions, encapsulated by functional correlation definitions, offering a way to conceptualize complex relationships in various contexts.

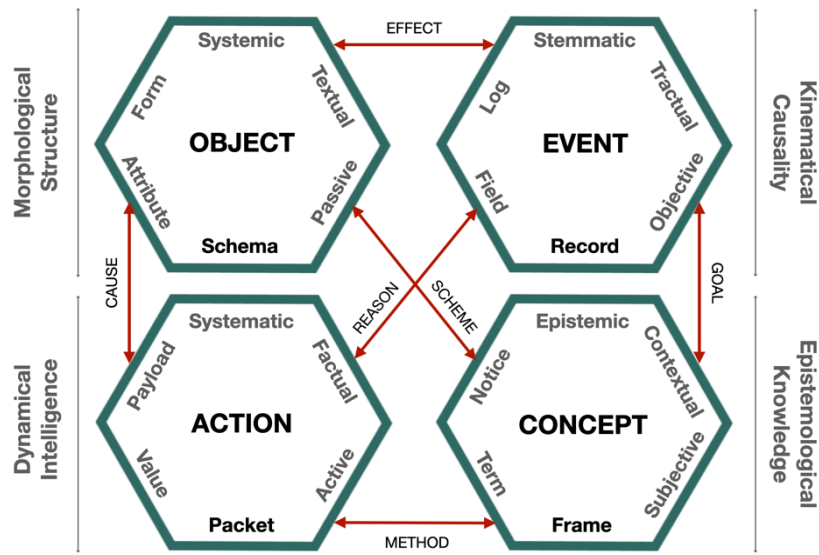

Figure 2. CDO functional correlations.

These are summarized in

Figure 2, and include:

- Scheme = Object + Concept
- Reason = Action + Event
- Cause = Action + Object
- Method = Action + Concept
- Goal = Concept + Event
- Effect = Object + Event

### 6.2.1 Scheme: Object + Concept

A scheme combines tangible elements ('Object') with abstract ideas ('Concept'). It involves practically applying concepts to physical entities and creating a structured plan or system. For instance, in architectural design (the scheme), the natural materials and space (objects) are shaped by aesthetic and functional ideas (concepts). This fusion results in a comprehensive, theoretically sound, and practically applicable blueprint.

### 6.2.2 Reason: Action + Event

A reason is the interplay between actions and their consequential events in this context. It involves understanding the rationale or justification linking what was done (action) to what subsequently happened (event). For instance, in scientific experiments, the reason for a specific result (event) can be attributed to the particular actions or procedures carried out. This perspective positions reason as the logical connection between actions and their effects, emphasizing how deliberate actions lead to specific outcomes or events.

### 6.2.3 Cause: Action + Object

A cause is the driving force behind an occurrence, originating from an action that directly impacts or changes an object. It represents the dynamic interaction between what is done (action) and what is affected (object). For instance, in everyday life, the cause of a broken window (object) might be traced to a child's ball hitting it (action). The focus here is on how a deliberate action influences a tangible entity, serving as the initial impetus for change.

### 6.2.4 Method: Action + Concept

A method is a procedural manifestation of a concept involving a series of actions. It's the practical implementation of an abstract idea or theory. Educational methodologies actively develop teaching strategies (methods) from educational theories (concepts) and execute them through specific actions like lecturing, interactive sessions, etc.

### 6.2.5 Goal: Concept + Event

A goal actively embodies an envisioned outcome (concept) and realizes it through significant occurrences or milestones (event). It starts as an idea or desired result and becomes tangible through key events marking progress or completion. In project management, completing a project phase (event) signifies attaining a part of the larger project goal (concept).

### 6.2.6 Effect: Object + Event

An effect is the outcome that arises from an interaction between an object and an event. It embodies an event's visible or measurable consequence concerning a specific object. For example, in medicine, the effect of a drug (event) on the human body (object) is observed through changes in health outcomes. In this framework, the effect signifies the manifestation of changes tied to and observed through events impacting objects.

## 7. ONTOLOGY INTRODUCTION

In the Informatics domain model, a supportive core data ontology is crucial in organizing and structuring contextual knowledge about the model within computational systems.

### 7.1. Informatics Domain Model

The purpose of developing a core data ontology for the Informatics domain model is to provide a unified and standardized approach to knowledge representation. The ontology facilitates efficient data integration, search, reasoning, and knowledge discovery by establishing a common understanding of the concepts, relationships, and semantics, acting as a foundation for data harmonization and interoperability in distributed data ecosystems.

The baseline ontology developed for the Informatics model follows a structured framework that aligns with the quadrimodal nature of the model. It encompasses the key concepts and relationships that define the core domains of objects, events, concepts, and actions. Each frame within the ontology represents a specific aspect or concept related to the modelled data, while the relationships capture the connections and dependencies between the ontological frames.

The core data ontology comprises four core data domains: objects, events, concepts, and actions. These domains provide interconnected relationships through design dependencies and associations. Here is a high-level textual representation of the ontology structure:

- **Objects** represent the structural aspects of data and its attributes. They define the morphological structure of systemic variables, data structures, functions, and methods.
- **Events** capture occurrences within the system, providing tractual fields of evidence for kinematical causality. These events help track the sequences and relationships between significant events or actions within the system, documenting the motion and causal connections, which are key to understanding how interconnected nodes relate.
- **Concepts** encompass the contextual terms, relationships, and knowledge of the data. They contribute to the epistemological knowledge of an epistemic concept or idea.
- **Actions** represent the operational and computational processes within the system, emphasizing the dynamical domain of system intelligence. These actions enable the system to execute transformations, processes, or functions on data, driving forward systematic operations and computational intelligence through action-driven operations.

Beyond the individual domains, cross-domain correlations show how the quadrimodal domains are interconnected and interdependent.

*"Consensual scheme"* represents an agreement by consensus on the objectual structure and conceptual interpretation of the data. It ensures a shared understanding and consistent interpretation of the data across computational systems.

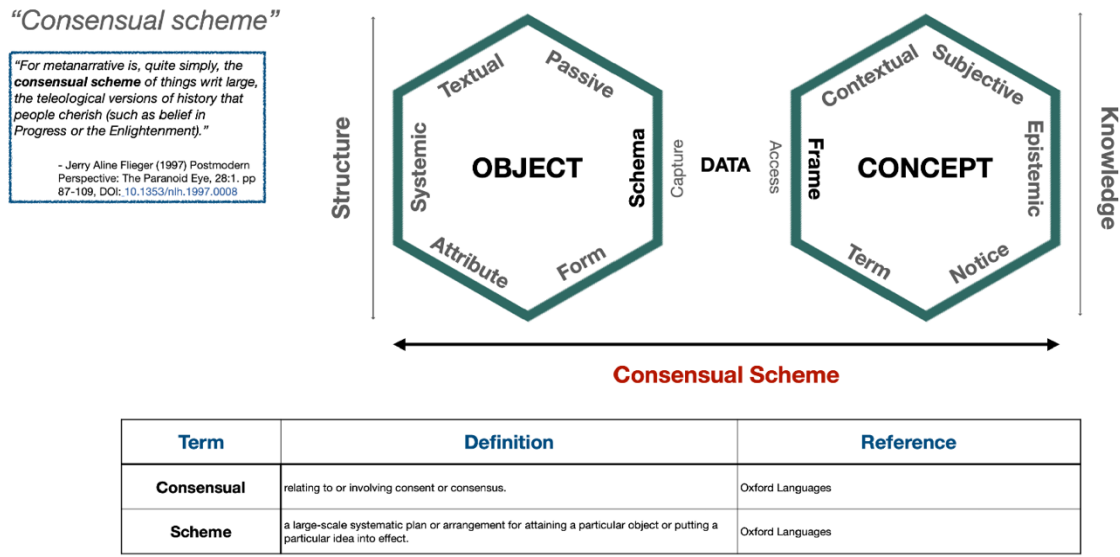

Figure 3. Consensual Scheme.

*"Sovereign reason"* emerges from the relationship between events and actions. It drives the behavior and functionality of the computational system, providing justifications and explanations for specific actions or events.

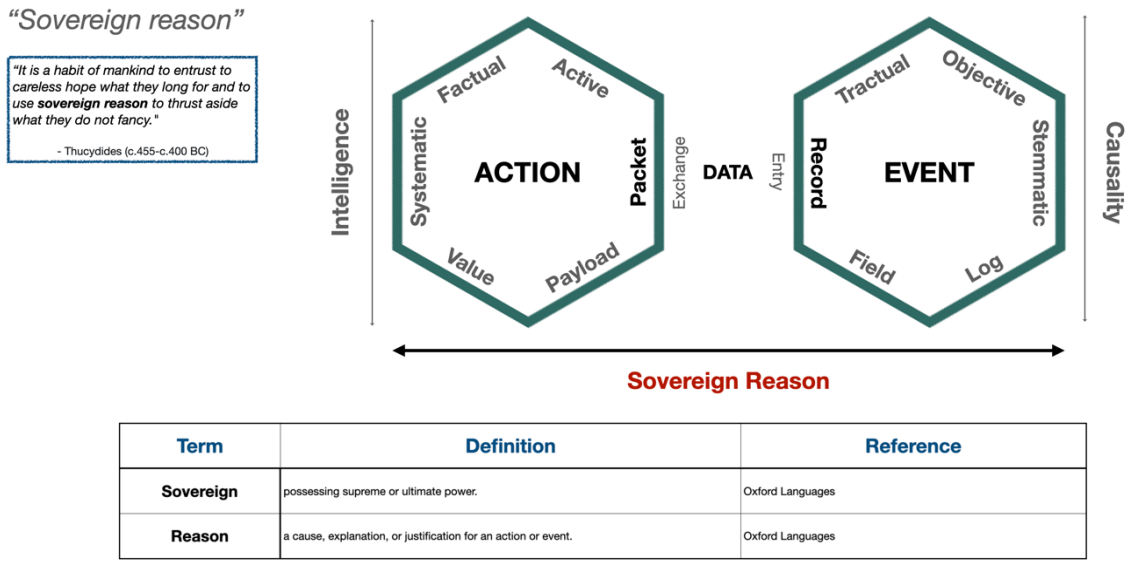

Figure 4. Sovereign Reason.

While this textual representation helps convey the structure of the ontology, it is essential to note that a visual graph representation provides a more intuitive and comprehensive understanding of the relationships and interconnections within the Informatics model.

The ontology[2] in Figure 5 is a graphical representation of the "Core Data Ontology," which visually represents the Informatics model to enhance its understandability and comprehensibility. This graph illustrates the relationships between concepts and highlights the hierarchy, dependencies, and interconnections within the Informatics model. The ontology provides a powerful visual tool for exploring the core data ontology, enabling users to navigate and comprehend the complex network of concepts and their relationships.

Key classes within the ontology include the representation of *Objects*, which define the structural aspects of data and its attributes. *Concepts* encompass the contextual terms, relationships, and knowledge associated with the data, fostering a shared understanding and facilitating effective communication. *Actions* drive dynamic processes, augmented intelligence, and computational operations that engage with the data, enabling efficient processing, analysis, and the extraction of meaningful insights. *Events* capture the occurrence of specific incidents or actions within the system, enabling a temporal understanding of the data flow.

The relationships within the ontology capture the dependencies and associations between classes, forming a comprehensive network of interconnections. The ontology introduces cross-domain correlations, such as the consensual scheme in Figure 3, between objects and concepts, establishing a standard agreement on the structure and interpretation of data. Similarly, sovereign reason in Figure 4 emerges from the relationship between events and actions, driving the behaviour and functionality of the computational system.

---

[2] Core Data Ontology repository represented as an OWL 2 ontology: Accessed at https://github.com/THCLab/ontology.

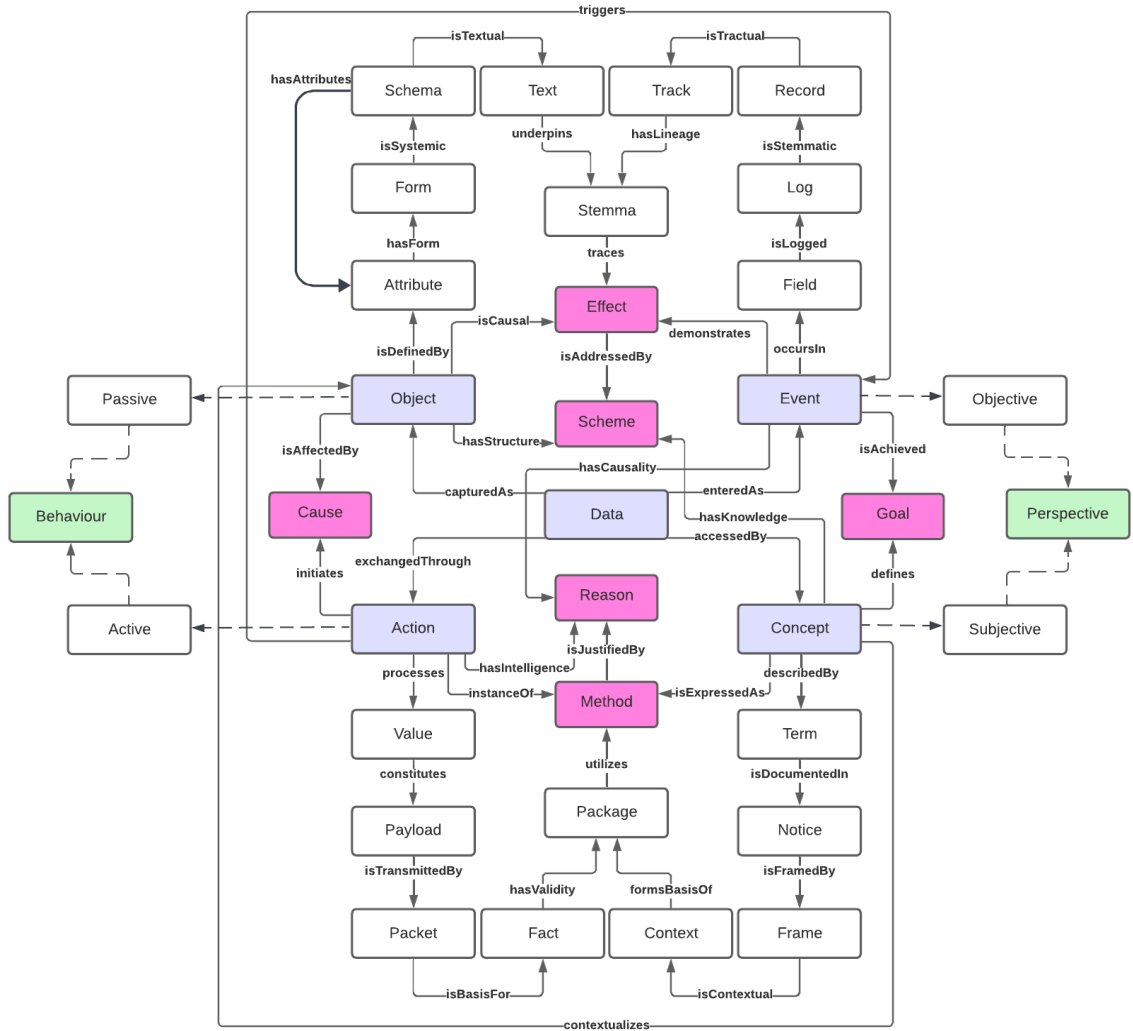

Figure 5. Core Data Ontology.

In summary, the core data ontology for the Informatics domain model plays a pivotal role in organizing and representing knowledge within computational systems. It provides a standardized framework for knowledge representation, fostering data interoperability, harmonization, and efficient processing. The knowledge graph visually presents the ontology, highlighting the key concepts and relationships and enabling users to navigate and comprehend the intricate knowledge of the Informatics model.

### 7.2. Ontology Engineering

The Core Data Ontology was designed using the ontology engineering methodology [1]. The methodology contains six key steps:

1. **Informal Competency Questions**: Gather requirements for the ontology in the form of informal competency questions, a set of use cases in natural language that a representational framework must be able to answer.
2. **Design**: Design the ontology using one or more design patterns [6], as discussed in the next section.
3. **Reuse**: Evaluate existing ontologies and reuse them when possible. For example, to implement time-related classes, reuse the time ontology [7].

4. **Implementation**: Implement the ontology in a formal language, such as OWL [8], in a way that captures all competency questions.
5. **Formal Competency Questions**: Represent the informal competency questions in natural language as formal knowledge graph queries. For example, competency questions for ontologies implemented in OWL can implemented with queries written in SPARQL [9].
6. **Evaluation**: The ontology is evaluated by testing each formal competency question in, say, SPARQL to ensure it returns the intended response from the knowledge graph.

In this paper, we have completed steps 1 through 4, with a caveat in step 3. While the ontology was designed to be reusable by vendors that utilize the CDO for sharing information, we did not reuse existing ontologies for its construction. This design decision ensures the CDO is a self-contained and stable model, with centralized control over the main classes, relationships, and properties. In future work, we will complete steps 5 and 6.

### 7.3. Ontology Design Pattern

The ontology was designed using a standard set of design patterns, namely the Ontology Design Patterns (ODPs). ODPs are conceptual tools that facilitate creating, managing, and reusing ontologies within information systems [6]. In the same way that ontologies are structured frameworks that encode a description of some aspect of the world, often to support tasks such as querying, searching, or integrating data [1], ODPs provide standardized solutions to recurring design problems, similar to design patterns in software engineering, which help ensure consistency and efficiency in developing ontologies. This chapter outlines the ontology patterns used to design the ODC ontology that captures the Informatics Model, outlining their role in improving the quality of ontology design by offering reusable components with explicit design rationales.

ODPs address several challenges associated with the complexity and reusability of ontologies. The lack of standard practices for reengineering and sharing expert knowledge about a domain results in large, monolithic structures that are difficult to reuse or adapt to new contexts [6]. By breaking down ontologies into smaller, task-oriented modules, ODPs aim to make ontology design more manageable, modular, and accessible. These patterns encapsulate best practices in ontology design, making it easier for designers to develop robust and reusable ontological frameworks.

### 7.4. CDO Ontology Design Pattern

The Informatics model plays several roles that connect the instances of *Object*, *Concept*, *Event*, and *Action* classes together through a shared data model. Firstly, ontologies created by vendor partners that reference CDO must be compatible vocabularies that are mapped to each other. Such mapping can be hierarchical if direct mapping is not available. CDO facilitates the integration of other ontologies using the *Object* and *Concept* classes. Secondly, events that modify states represented by CDO and partner vendors must be tractable and reproducible as events in the CDO logs. These are covered in the following subsections.

#### 7.4.1. Object and Concept Mapping Pattern

Two design patterns are candidates that fit the first requirement: the mapping ontology pattern and the content ontology pattern [6]. The mapping ontology pattern facilitates the following requirements:

- Encodes conceptual ontology patterns, not logical axioms. Supported relationships:

- o   equivalence, containment, overlap
- o   not equivalent, not contained, not overlap or disjoint

However, these are too strong for the ODC use cases. Hence, the content ontology pattern was chosen. It facilitates the following requirements:

- Ontology reuse through specialization.
- Ontology reuse through extension.

Specifically, local concepts were created that follow the linguistically relevant components. They frame a concept in a way that can be expressed linguistically, grounded in the role each concept plays. Here, the mapping (or alignment) does not require composition. Rather, mapping between terms is sufficient.

The *Object* and *Concept* classes facilitate representing local ontology compatible with CDO to identify which components are mappable. Here, the *Object* instance is the reference point for representing "things" in the real world. The *Concept* class facilitates mapping between external *Concepts* to a single *Object* instance. For example, CDO can map between languages using content ontology design patterns such as:

    :apple <=> :pomme
    :apple <=> :apfle

where :apple is an instance of the Object class, while the French :pomme and German :apfel are instances of the *Concept* class mapped to :apple.

Three types of matching are supported by the content pattern:

1. **Broader Matching**: Used when mapping a more general local *Concept* (e.g. :fruit) to a specialized *Object* (e.g. :apple). It may need to be specialized to address the specific domain requirements after being mapped.
2. **Narrower Matching**: Used when mapping a more specific local *Concept* (e.g. :fuji_apple) to a general *Object (e.g ;apple)*. It may need to be generalized to fit the broader requirements of the local domain.
3. **Partial Matching**: Used when an *Object* only partially matches a local use case. The use case should be divided into smaller parts, and multiple CPs may need to be selected and combined to cover all aspects. Expansion of the CPs may be necessary in all scenarios.

**7.4.2. Activity and Event Time Indexed Pattern**

A time-indexed content pattern is a specialized content pattern that captures changes in ontological instances, similar to the work by Presetti and Gangemi (2008) [11]. The *Action* class facilitates the transformation of data and ontological definitions, with distinct times when changes occur. The *Event* class captures state changes that track occurrences of activities as logged events. For example, when generalizing a *Concept* from :fuji_apple to :apple, the event would be logged with the previous state (:fuji_apple), final state (:apple), and a timestamp of the event.

## 8. APPLICATIONS AND BENEFITS

The Informatics Domain Model and its accompanying Core Data Ontology (CDO) offer extensive applicability across various domains. By structuring data into objects, events, concepts, and actions within the CDO framework, organizations can leverage these capabilities for better data management, intelligent decision-making, and legal compliance. We outline several practical examples below, showing how this model applies to real-world scenarios, particularly AI auditing, robotics, and legal frameworks, and its exciting potential for these fields.

### 8.1 AI Auditing and Consent Management

The increasing use of AI systems in various industries has raised significant concerns about the ethical use of data. Securing user consent to include personal data in AI training sets is becoming a critical requirement from a compliance perspective and ensuring the auditability and transparency of AI models. The Core Data Ontology facilitates the management of consent data, providing a structured framework that enables organizations to track, verify, and audit the use of personal information in AI models.

By integrating consent receipts into the data lifecycle, organizations can maintain a traceable record of user approvals, ensuring that AI systems only use authorized data. This approach enhances trust and compliance with privacy regulations such as GDPR. The CDO's semantic layer can categorize and track different consent statuses, allowing for dynamic enforcement of data usage policies based on user preferences. Auditing the training data of AI models through the CDO framework ensures ethical sourcing of all data, reducing the existential risks associated with unauthorized AI decision-making.

### 8.2 Robotics and Multimodal AI

The Informatics Domain Model, through the CDO, holds significant potential for advancing multimodal AI and robotics by offering a sophisticated framework for data categorization and interaction. In particular, the model's ability to integrate semantic mapping and classification for real-world objects can revolutionize how robotics systems interpret and interact with their environments.

For example, the concepts and objects defined in the CDO can be enriched through computer vision and other AI-based sensory modalities, allowing robotics systems to perceive and classify physical objects more effectively. The CDO provides the foundational tool for labeling and classifying objects based on their material properties, semantic relationships, and functional purposes. This capability enables robots to perform more accurate multimodal classification, improving their ability to recognize, interact with, and respond to complex real-world scenarios.

As large vision-language models evolve, the CDO will facilitate visual data fusion with linguistic context, supporting advanced robotic reasoning capabilities. By mapping physical objects to their conceptual representations, robotics systems can achieve better situational awareness, improving decision-making and autonomy in various applications, such as industrial automation or healthcare.

### 8.3 Data Provenance and Legal Implications

In a data-driven world, ensuring the provenance of digital assets is becoming increasingly important, particularly in legal and intellectual property (IP) contexts. The Core Data Ontology is crucial in establishing a verifiable chain of data custody, ensuring that data's origin, modifications, and usage are recorded and traceable. This capability is vital for proving the authenticity of data in legal disputes, intellectual property claims, and compliance audits.

For instance, digital provenance will become a verifiable legal precedent, with courts relying on systems like the CDO to establish who created, modified, or used specific datasets. The ability to link events with particular actions and objects, as provided by the CDO, ensures that trackers can reliably trace data back to its source. Organizations can use this framework to enforce data integrity and demonstrate ownership over proprietary information, safeguarding intellectual property rights and minimizing legal risks.

As data provenance becomes a core issue in technology and innovation, the CDO provides a scalable and adaptable framework to meet these evolving demands. By integrating digital provenance directly into the data management lifecycle, organizations can ensure that their data assets maintain legal integrity, even as they scale across distributed systems and multi-stakeholder environments.

### 8.4. Knowledge Management and Discovery

The Informatics model facilitates effective knowledge management and discovery by providing a structured framework for organizing and categorizing knowledge assets. By leveraging the concepts domain and the relationships within the ontology, organizations can create knowledge graphs, semantic networks, and intelligent search systems that enable efficient knowledge discovery, information retrieval, and knowledge sharing, promoting enhanced collaboration, accelerating innovation, and fostering continuous learning within organizations.

For example, a research institution can utilize the Informatics model to manage and organize its vast collection of research papers, publications, and knowledge assets. The institution can develop a robust knowledge discovery platform by categorizing and linking concepts, creating relationships between authors and topics, and leveraging the ontology's vocabulary. Researchers can easily navigate the knowledge base, uncover related information, and gain valuable insights for their research projects.

These practical examples demonstrate the applicability and transformative potential of the Informatics Domain Model and the Core Data Ontology and underline their role in addressing critical challenges in AI, robotics, and data governance. Organizations can confidently tackle these challenges by adopting this data-centric approach, setting the stage for future advancements in data-driven innovation and legal frameworks.

## 9. SCALABILITY AND FUTURE DIRECTIONS

The meticulous design of the Informatics domain model and its accompanying ontology provide a scalable and adaptable foundation for computational systems. The accuracy and precision of the definitions, sourced primarily from the Oxford Dictionary, ensure a solid basis for the model's quadrimodal structure. Like a well-constructed jigsaw puzzle, the natural synergy of these definitions seamlessly meshes to form a cohesive and interrelated framework. This precise and cohesive foundation instills confidence in the model as the bedrock of any computational system, fostering trust and reliability in its applications.

The Informatics domain model facilitates the communication and operationalization of knowledge transformation in a scalable distributed peer-to-peer system. As Knowles et al. point out [12], to ensure harmonization between such systems at scale, a model for decentralized semantics must be present. This is a departure from the long history of research that focused on the interoperability of services in cloud-based ecosystems, specifically, the configuration, discovery, and interactions between microservices [13]. Related work includes the Overlays Capture Architecture (OCA) [14], an explicit representation of task-specific objects with deterministic relationships to other objects while also providing framing mechanisms into epistemological concepts. Another example is the LinkML framework, which uses a predefined linked data modelling language to align concepts across systems, represented as OWL ontologies, SHACL rules, and the LinkML Schema [15]. Several "schema" representations exist and have been mapped automatically based on various characteristics [16]. Others still have focused on the data portion, namely distributed systems of "big data" stores, to share information between systems [17].

The robust nature of the Informatics model's definitional accuracy offers assurance of its long-term scalability. Organizations can confidently scale their systems and leverage the model's intrinsic design, which separates the semantic-based domains of "consensual scheme" (morphological and epistemological semantics) and the mechanic-based domains of "sovereign reason" (kinematical and dynamical mechanics). The Informatics model can seamlessly integrate with existing and future ontologies and vocabularies, serving as a stable root from which to seed any knowledge concepts. This adaptability and compatibility make the Informatics model invaluable for expanding and evolving computational systems across various domains.

We do not anticipate significant updates or revisions to the Informatics domain model anytime soon. The rigorous research process and the accuracy of the definitional synergy throughout the model have resulted in a robust and comprehensive framework. However, as with the maturing of a mighty oak tree, the core data ontology may naturally evolve to include thematic nuances that further enrich the knowledge graph. These incremental enhancements will build upon the existing foundation, adding new layers of depth and specificity to accommodate emerging requirements and expanding domains.

Ongoing and future research endeavours will focus on exploring the practical applications of the Informatics model in diverse fields such as healthcare, finance, education, and more. Directed efforts will leverage the model's capabilities for advanced knowledge management, data-driven decision-making, and semantic interoperability. Additionally, research will continue to investigate integrating the Informatics model with emerging technologies, including artificial intelligence, decentralized technologies, and the Internet of Things, to unlock new possibilities and propel the growth of intelligent and interconnected systems. The Core Data Ontology will be fully evaluated by the ontology engineering methodology described in section 7. Specifically, competency questions will be formalized, and a knowledge graph of object, concept, action, and event instances will be created. Formal competency questions will then be used to evaluate whether the ontology satisfies requirements for specific application domains, as outlined in section 8.

The scalability and future directions of the Informatics model are grounded in its robust architecture, definitional accuracy, and adaptability. By embracing the model as a foundational underpinning, organizations can confidently navigate the ever-evolving landscape of computational systems, harnessing the power of semantic design and knowledge representation to drive innovation, efficiency, and meaningful insights.

## 10. CONCLUSION

In this paper, we introduce the Informatics domain model[3], a novel paradigm for computational systems that places data at the core of decision-making and security. The model's quadrimodal structure, encompassing objects, events, concepts, and actions, provides a comprehensive framework for data-centric design, enabling precise categorization and granular control over data elements. The Informatics model offers a new perspective on securing and managing data in the digital age by shifting the focus from node-centric approaches to data-centric design principles.

---

[3] We invite readers to refer to the "Informatics Domain Model" presentation https://zenodo.org/records/13729820 for a more visual representation of the model, offering a more comprehensive view of the model's core domains, relationships, and vocabulary definitions, aiding in its understanding and application.

Throughout this paper, we have highlighted the challenges and limitations of current informatics models, emphasizing the need for a more comprehensive and scalable solution. The Informatics model addresses these challenges by offering a robust foundation that combines definitional accuracy, semantic design, and knowledge representation. By leveraging the model's design principles of consensual scheme and sovereign reason, organizations can achieve enhanced data security, efficient role-based access control, and semantic interoperability.

The significance of the Informatics domain model and core data ontology lies in its ability to provide a common language and framework for diverse domains and industries. The model's accuracy, derived from meticulous research and using trusted sources such as the Oxford Dictionary, instills confidence in its definitions and structures. Based on the Informatics domain model, the underlying ontology is a stable and adaptable root for categorizing and understanding knowledge concepts, facilitating effective communication and collaboration.

In conclusion, the Informatics domain model and supporting core data ontology present a transformative approach to computational systems, enabling secure, efficient, and meaningful data management. By adopting the Informatics model, organizations can unlock the full potential of data-centric design, fostering innovation, interoperability, and informed decision-making. We encourage researchers, practitioners, and stakeholders to embrace the Informatics model and explore its possibilities, contribute to its further development, and shape the future of computational systems.

## ACKNOWLEDGEMENTS

I want to thank all the participants and contributors from the Decentralized Semantics Working Group at the Human Colossus Foundation for their invaluable input. Special thanks to Mark Lizar (0PN), Carly Huitema (University of Guelph), Neil Thomson (QueryVision), Scott Warner (Secours.ai), Steven Milstein (Collab.Ventures), and Ryan Barrett (Stratosoft), whose expertise and collaboration have greatly enriched this work. A special mention goes to Bart Gajderowicz, PhD, ontology engineer from the University of Toronto, and Keith Dugas, augmented intelligence expert from Secours.ai, for their significant contributions and insights that helped shape this paper. Finally, I appreciate the countless others who have provided contributions and insights on decentralized semantics over the past four years. Your support has been instrumental in bringing this work to fruition.

**Authors**

**Paul Knowles** is the inventor of 'Decentralised Semantics' and co-founder of the Human Colossus Foundation, where he serves as Chair of the Decentralised Semantics Working Group. With over 25 years of experience in pharmaceutical biometrics at companies such as Roche, Novartis, GlaxoSmithKline, Amgen, and Pfizer, he has become a leading figure in developing decentralised data infrastructures. His contributions include the Overlays Capture Architecture (OCA), a framework for semantic interoperability, and he holds advisory roles at Secours.ai and in Global Privacy Rights, 0PN Governance Architecture, Transparency Lab, and Standards Engineering at 0PN. Paul is a published author on self-sovereign identity and decentralised semantics, with papers focusing on distributed health data ecosystems and harmonising quantitative medical data.

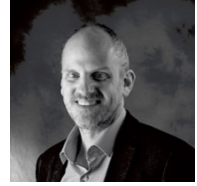

**Bart Gajderowicz** has a BSc and MSc in Computer Science and a PhD in Industrial Engineering. He is a Senior Research Associate at the Industrial Engineering Department and Executive Director at the Urban Data Centre at the School of Cities, University of Toronto. He is a technical advisor on topics ranging from artificial intelligence, impact measurement and ontology engineering for multiple organizations. He manages the Canadian Urban Data Catalogue project (CUDR), is a co-author of the Common Impact Data Standard (CIDS), and the Dataset Metadata Capability Maturity Model (DMCMM) series of standards, the lead researcher on SMILE, an AI model for the opportunistic reasoning about information in a knowledge graph and selecting knowledge sources for information extraction from text, and TIMM, a traceable impact model for supply chains. He was the director of the BRAMA project, a high-fidelity simulation environment for emotion-based reasoning of social services clients and an ontology of social service needs (OSSN).

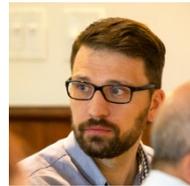

**Keith Dugas** is a seasoned executive and thought leader in digital mobility and artificial intelligence. He serves as Director of Operations AI at Air Canada and Chief Technology Advisor at Secours.ai. A sought-after keynote speaker, Keith has presented at prominent industry events, including AGIFORS, AeroID, Aviation TechWeek, and IATA MCC. He holds a bachelor's degree in Sociology and Psychology from the University of Ottawa, along with certificates from UC Berkeley and MIT.

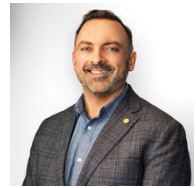